\documentclass[sigconf]{acmart}

\usepackage{amsmath}
\usepackage{xcolor}
\usepackage{makecell}
\usepackage{framed}

\AtBeginDocument{%
  }

\copyrightyear{2026}
\acmYear{2026}
\setcopyright{cc}
\setcctype{by}
\acmConference[EASE '26]{International Conference on Evaluation and Assessment in Software Engineering}{June 09--12, 2026}{Glasgow, United Kingdom}
\acmBooktitle{International Conference on Evaluation and Assessment in Software Engineering (EASE '26), June 09--12, 2026, Glasgow, United Kingdom}
\acmDOI{10.1145/3816483.3816519}
\acmISBN{979-8-4007-2348-3/2026/06}

\begin{document}

\title[Detecting and Resolving Pragmatic Ambiguities in NLRs]{A Retrieval-Augmented Framework for Detecting and Resolving Pragmatic Ambiguities in Natural Language Requirements}

\author{Pavithra PM Nair}
\affiliation{%
  \institution{Tata Consultancy Services}
  \city{Pune}
  \country{India}
}
\email{pavithra.nair@tcs.com}

\author{Preethu Rose Anish}
\affiliation{%
  \institution{Tata Consultancy Services}
  \city{Pune}
  \country{India}
}
\email{preethu.rose@tcs.com}
\renewcommand{\shortauthors}{Nair and Anish}

\begin{abstract}
Natural language requirements (NLRs) are essential for bridging communication gaps among diverse stakeholders in software development. However, the inherent ambiguity in NLRs can pose significant challenges. In particular, some requirements may be misinterpreted due to varying contextual knowledge and domain-specific expectations of the stakeholders, a phenomenon known as pragmatic ambiguity. This paper presents an approach for detecting and resolving pragmatic ambiguities in NLRs. The approach leverages retrieval-augmented generation techniques with \textit{novice}, \textit{intermediate}, and \textit{expert} domain knowledge bases to simulate stakeholders with varying domain expertise and detect discrepancies in requirement interpretation. Candidate disambiguated requirements are generated using the \textit{expert} domain knowledge base, with final validation by a requirements analyst required to ensure alignment with the intended system functionality. We evaluate the approach on two requirements specification documents from the PUblic REquirements (PURE) dataset, using four large language models: GPT-4o-mini, Mistral-7B, Llama-3.1-8B, and Qwen2.5-7B. Detection performance is assessed using macro-averaged accuracy, precision, recall, F1, and F2 scores. The resolution quality of the candidate disambiguated requirements is measured through human evaluation of relevance, clarity, and consistency. In this initial evaluation, results show that the proposed approach can detect pragmatic ambiguities and produce candidate disambiguated requirements that are relevant, clear, and consistent with the intended system functionality. Among the evaluated models, GPT-4o-mini achieved the highest macro-averaged recall (0.75) and F2 score (0.75) for pragmatic ambiguity detection. In the resolution task, GPT-4o-mini received the highest relevance scores from human evaluators, while Mistral-7B achieved the highest scores for clarity and consistency.
\end{abstract}

\ccsdesc[500]{Software and its engineering~Requirements analysis}

\keywords{Pragmatic Ambiguity, Requirements Engineering, Natural Language Requirements, Large Language Models, Retrieval-Augmented Generation}


\maketitle

\section{Introduction}
Natural language requirements (NLRs) are a crucial element in the software development life cycle (SDLC), as they capture both the functional specifications and non-functional attributes of the software to be developed. Serving as a communication bridge between different stakeholders in the SDLC, NLRs guide the design, implementation, and validation of software systems. However, the inherent flexibility and context-dependence of natural language often introduce ambiguities in these requirements \cite{berry2003ambiguity, massey}. NLRs may be interpreted differently depending on each stakeholder’s background, experience, and perspective \cite{Ferrari2019AnNA}.

Misinterpretations can arise not only between requirements analysts (RAs) and domain experts, but also among stakeholders from different domains or within stakeholder groups with varying levels of expertise. During requirements elicitation meetings, RAs and domain experts often use specialized jargon \cite{Tubo2011SystematicRA, Zowghi2005}. Differences in domain knowledge and terminology between participants can create misunderstandings, leading to unclear or underspecified requirements \cite{christel_1992}. Such ambiguities may propagate through subsequent development phases, leading to costly rework in later stages of the SDLC \cite{boehm2009software, davis1993identifying}.

Consider the requirement: \textit{``The system must provide real-time analytics on server performance."} For a system administrator, ``real-time" may imply the need for monitoring tools that update data at short intervals, such as every minute, to ensure optimal server performance. Conversely, a software engineer may interpret ``real-time" as requiring millisecond-level precision, potentially involving integration with lower-level hardware sensors for high granularity monitoring. Meanwhile, a business stakeholder might understand ``real-time" as the need for data updates only at a frequency sufficient for business decision-making, which could range from hourly to daily intervals.

As illustrated by this example, the same term, ``real-time," can be interpreted in vastly different ways depending on the domain knowledge and priorities of the involved stakeholders. This type of ambiguity, arising from differing contextual knowledge and domain-specific expectations, is referred to as pragmatic ambiguity \cite{collective, berry2003ambiguity}. 

Pragmatic ambiguities in NLRs have been relatively underexplored, with a limited body of research addressing them by incorporating domain knowledge into requirements analysis \cite{ Ferrari2019AnNA, 6894849}. However, challenges remain, particularly in capturing all potential pragmatic interpretations of a requirement. The diverse backgrounds, knowledge, and expertise of stakeholders make it difficult to develop infallible rule-based systems for detecting such ambiguities. While it may be infeasible to create a tool that accounts for every possible interpretation, we believe that RAs could significantly benefit from recommender systems designed to assist in resolving pragmatic ambiguities during the requirements analysis process.

In this paper, we propose a novel framework that detects pragmatic ambiguities in NLRs. Our approach identifies potentially ambiguous terms in requirements and generates elucidation questions (EQs) --- targeted, open-ended questions tied to a specific ambiguous term in a requirement, designed to surface the clarification needed to resolve its pragmatic ambiguity. It leverages retrieval-augmented generation (RAG) techniques, incorporating \textit{novice}, \textit{intermediate}, and \textit{expert} domain knowledge bases, which represent varying levels of stakeholder domain knowledge. Using these knowledge bases, we simulate how stakeholders with differing levels of domain expertise would interpret requirements, enabling our framework to detect discrepancies in interpretation. Additionally, we leverage the \textit{expert} domain knowledge base to propose candidate disambiguated requirements (referred to as ``resolutions'') for the detected pragmatic ambiguities. These candidate resolutions reflect the understanding and implicit assumptions held by domain experts; however, final validation by an RA is essential to ensure that the candidate resolutions align with intended system functionality. 

The framework is intended for use during the requirements analysis phase, after requirements elicitation and before system development. It is designed to surface ambiguities in requirement interpretation between RAs and domain experts, as well as among stakeholders who have varying levels of domain expertise, all of which the RA should consider when analyzing requirements.

We structure our study around the following research questions:
\begin{itemize}
    \item \textbf{RQ\textsubscript{1}: }\textit{How effectively can pragmatic ambiguities in NLRs be detected using a retrieval-augmented method that simulates varying levels of stakeholder domain expertise?}
    \item \textbf{RQ\textsubscript{2}: }\textit{To what extent are the candidate resolutions relevant, clear, and consistent with the intended system functionality?}
\end{itemize}

To answer these questions, we conduct an empirical evaluation using two requirements specification documents from the publicly available PURE dataset \cite{Ferrari2017PUREAD} and four large language models (LLMs) (GPT-4o-mini \cite{openai2024gpt4omini}, Llama-3.1-8B \cite{meta2024llama31}, Mistral-7B \cite{mistral2023mistral7b}, and Qwen2.5-7B \cite{qwenlm2024qwen25}). Detection performance is assessed using macro-averaged accuracy, precision, recall, F1, and F2 scores. Candidate resolutions are evaluated using three human evaluation metrics: \textit{Relevance}, \textit{Clarity}, and \textit{Consistency}. Among the models assessed, GPT-4o-mini achieved the highest macro-averaged recall (0.75) for pragmatic ambiguity detection, along with a strong macro-averaged F2 score (0.75). GPT-4o-mini achieved the highest \textit{Relevance} score, while Mistral-7B outperformed the other models on \textit{Clarity} and \textit{Consistency} metrics.

The main contributions of this paper are as follows:
\begin{itemize}
    \item We propose a framework for detecting pragmatic ambiguities in NLRs by generating EQs and comparing interpretations across \textit{novice}, \textit{intermediate}, and \textit{expert} domain knowledge bases.
    \item We explore an approach for addressing detected pragmatic ambiguities by generating candidate resolutions informed by the understanding and implicit assumptions captured in the \textit{expert} domain knowledge base.
    \item We conduct an empirical evaluation on two requirements specification documents from the PURE dataset, demonstrating the framework’s effectiveness using both quantitative detection metrics and human evaluation of resolution quality.
    \item We release a replication package containing the ground truth datasets, code, and experimental outputs, to support reproducibility and future research\footnote{\url{https://github.com/pavithranair/Pragmatica}}.
\end{itemize}

The remainder of this paper is structured as follows. Section~\ref{background} reviews related work on ambiguity in requirements. Section~\ref{approach} describes our proposed approach. Section~\ref{experimental_setup} presents the experimental setup. Section~\ref{results} reports our findings. Section~\ref{threats} discusses threats to validity, and Section~\ref{conclusion} concludes the paper.

\section{Background}
\label{background}
This section provides background on ambiguity in NLRs and reviews related work in the area. Section~\ref{classification} classifies the types of ambiguities commonly found in NLRs, Section~\ref{ambiguity} surveys existing approaches to address these ambiguities, and Section~\ref{pragmatic} focuses on prior research specifically addressing pragmatic ambiguity.
\subsection{Classification of Ambiguity in NLRs}
\label{classification}
Ambiguities in NLRs are generally classified into four primary categories, as outlined by Berry et al. \cite{berry2003ambiguity}:
\begin{itemize}
    \item Lexical ambiguity - arises when a term has multiple meanings.
    \item Syntactic ambiguity - occurs when a sentence can be parsed into multiple grammatical structures, each with a different meaning.
    \item Semantic ambiguity - stems from a sentence that can be interpreted into more than one logical expression.
    \item Pragmatic ambiguity - arises when a sentence has multiple interpretations depending on the context in which it is used.
\end{itemize}

For pragmatic ambiguity, the concept of ``context'' is broadly defined by Ferrari and Gnesi \cite{collective}, encompassing multiple levels: (1) the requirements immediately preceding and following the current one, (2) other requirements in different sections of the document, (3) the domain knowledge of the stakeholder reading the requirement, and (4) the stakeholder’s common sense knowledge. 

In addition to these core ambiguity types, the literature highlights two phenomena that are closely related to ambiguity in the context of NLRs:
\begin{itemize}
    \item Vagueness - occurs when an expression has borderline cases, where its truth value is indeterminate, typically due to the use of adjectives or adverbs.
    \item Generality - involves expressions that are overly broad and require further specification to convey a precise meaning \cite{cruse1986lexical}.
\end{itemize}
Examples of these ambiguity categories and related phenomena are detailed in the work of Berry et al. \cite{berry2003ambiguity}.

Massey et al.~\cite{massey} proposed a more recent taxonomy that includes 
lexical, syntactic, semantic, vagueness, referential, and incompleteness 
ambiguity, treating vagueness as a type of ambiguity and subsuming aspects 
of pragmatic ambiguity under referential ambiguity.

\subsection{Approaches to Address Ambiguities in NLRs}
\label{ambiguity}

Existing techniques for ambiguity detection in NLRs are primarily rule-based, relying on linguistic patterns to identify ambiguities within requirements. Tools such as QuARS \cite{Lami2019} and SREE \cite{10.1007/978-3-642-37422-7_6} exemplify this approach, which has matured to the point of being applied in industrial projects \cite{FEMMER2017190, Ferrari2018DetectingRD}. These tools identify ambiguous phrases based on linguistic markers and typically require manual review by RAs. Commercial tools, such as Qualicen Scout \cite{Femmer2018RequirementsQD}, also use rule-based methods to analyze NLRs for ambiguity. However, rule-based approaches tend to struggle with detecting semantic and pragmatic ambiguities, as they primarily address lexical and syntactic issues and lack scalability for more complex ambiguity types.

Statistical techniques leverage machine learning and statistical models to 
detect ambiguities. Ezzini et al.~\cite{9793957} employed supervised 
classifiers and SpanBERT~\cite{joshi-etal-2020-spanbert} to detect anaphoric 
ambiguities in requirements. Chantree et al.~\cite{chantree} addressed 
coordination ambiguities using word distribution heuristics. While statistical 
techniques have advanced, like rule-based approaches, they often focus on 
lexical and syntactic ambiguities.



\subsection{Approaches to Detect and Resolve Pragmatic Ambiguity in NLRs}
\label{pragmatic}
Several studies have made notable contributions to addressing pragmatic ambiguity in NLRs. Ferrari et al. \cite{6894849} proposed a method for detecting pragmatic ambiguities by modeling the background knowledge of different stakeholders using graph-based representations. Their approach employs a shortest-path search algorithm to simulate varying pragmatic interpretations of a requirement. By comparing these interpretations, this method identifies potential ambiguities. However, this approach can be computationally expensive, as it requires the construction and traversal of multiple domain knowledge graphs for each requirement, which becomes particularly resource-intensive in large or complex domains with many stakeholders. 

In subsequent work, Ferrari et al. \cite{Ferrari2019AnNA} introduced an NLP-based approach for detecting cross-domain ambiguities in requirements engineering. They build domain-specific language models using corpora for each domain, derive word embeddings, and compare usage neighborhoods of terms across domains to estimate ambiguity. They rank terms by ambiguity score and validate against human annotations. While their work is powerful for detecting terms likely to cause misunderstanding between different domains, it does not address ambiguity within a domain caused by varying levels of expertise among stakeholders. 

Tools such as MaRK \cite{mark}, WIKINA \cite{wikina}, and WikiDoMiner \cite{wikidominer} provide mechanisms for retrieving domain-specific knowledge that can support stakeholder understanding. However, these tools do not explicitly detect or resolve pragmatic ambiguities, and significant manual effort remains necessary to identify and clarify ambiguous requirements.

Our work builds on Ferrari et al.’s approaches \cite{6894849, Ferrari2019AnNA}. Our approach leverages multiple knowledge bases that simulate varying levels of domain expertise within the same domain. Cross-domain ambiguities often stem from incomplete knowledge about the domain of the system for which the requirements are specified, and even when stakeholders appear to have different priorities, these differences often trace back to variations in understanding the system’s capabilities, constraints, or context \cite{jain2020crossdomainambiguitydetectionusing, Ferrari2019AnNA}. In our approach, cross-domain stakeholders may be considered as novices in the system's domain, allowing the framework to naturally capture both cases: ambiguities that arise within the system’s domain due to varying levels of domain expertise, as well as certain ambiguities that surface across domains. In this way, our work generalizes and extends prior approaches.

We do not provide a direct empirical comparison with prior approaches to pragmatic ambiguity detection, as such a comparison is not technically feasible without compromising methodological validity. Ferrari et al.'s graph-based approach does not provide a publicly released dataset corresponding to the original study, making faithful reproduction of the experimental setting infeasible \cite{6894849}. Although implementations and resources are available for the later cross-domain approach \cite{Ferrari2019AnNA}, this approach detects ambiguity at the term level. Aligning its outputs with requirement-level pragmatic ambiguity detection would require additional design choices, related to corpus selection and aggregation of term-level signals, which can substantially affect ambiguity outcomes and confound comparative evaluation. Accordingly, we position our contribution as complementary to prior work.

\section{Proposed Approach}
\label{approach}

\begin{figure}[h] 
    \centering
    \includegraphics[width=0.9\columnwidth]{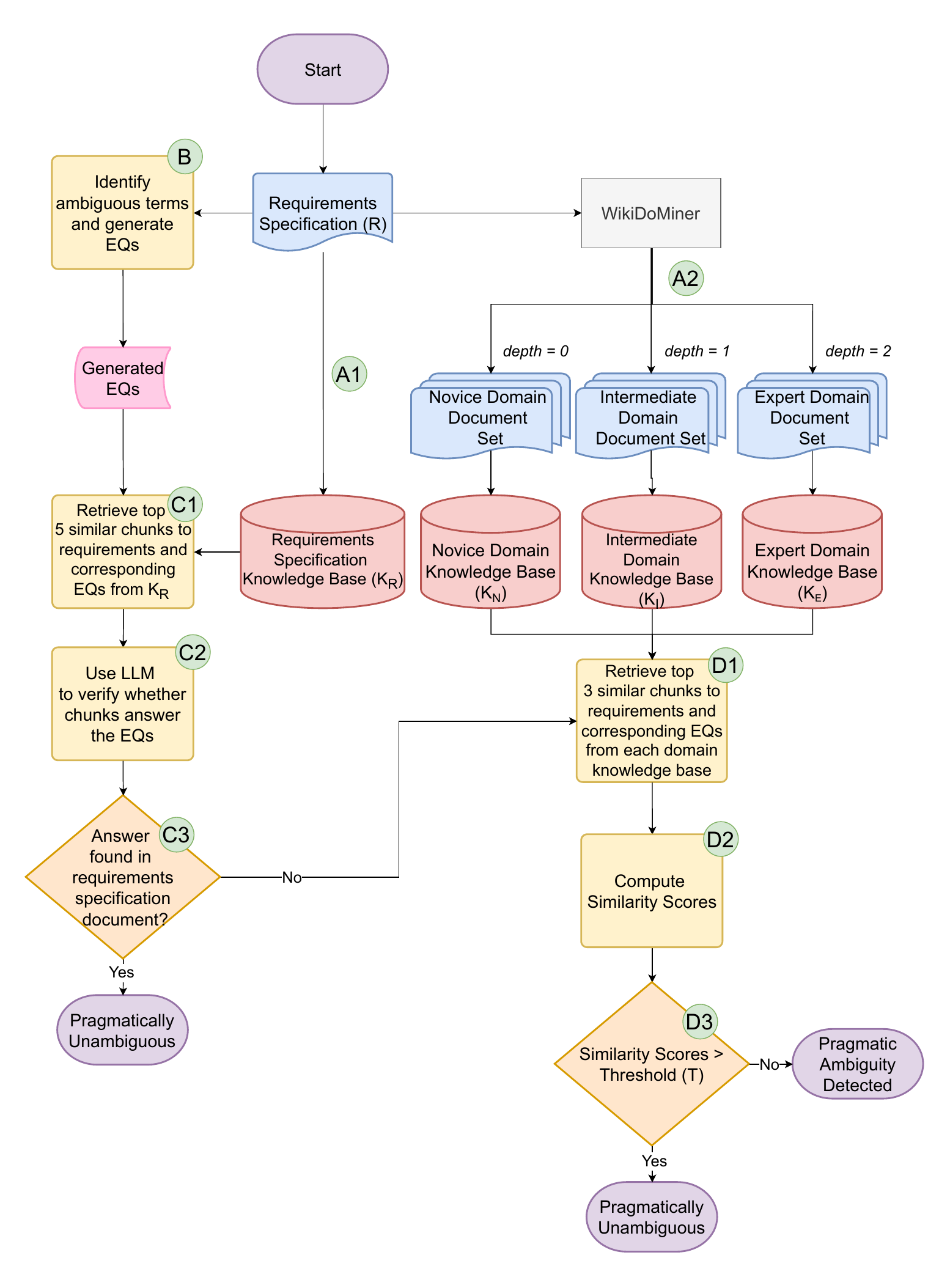} 
    \caption{Proposed Approach for Pragmatic Ambiguity Detection}
    \label{fig:detection}
\end{figure}

In this section, we outline our approach for pragmatic ambiguity detection 
and resolution, illustrated in Figure~\ref{fig:detection} and 
Figure~\ref{fig:resolution}, respectively. In the detection phase, a 
requirements specification knowledge base (\(K_R\)) is constructed from 
the input document, and three domain knowledge bases (\(K_N\), \(K_I\), 
\(K_E\)) are built using WikiDoMiner~\cite{wikidominer} to simulate novice, 
intermediate, and expert levels of domain expertise. An LLM generates EQs 
for each requirement to surface potentially ambiguous terms. Each requirement 
and its EQs are first queried against \(K_R\); if the specification resolves 
the ambiguity, the requirement is treated as unambiguous. Otherwise, the 
same query is posed to \(K_N\), \(K_I\), and \(K_E\), and the semantic 
similarity between the retrieved interpretations is computed. If similarity 
falls below a defined threshold, the requirement is flagged as pragmatically 
ambiguous. In the resolution phase, candidate resolutions are generated by 
querying the requirement and its EQs against \(K_R\) and \(K_E\), and are 
subsequently validated by an RA to ensure alignment with the intended system 
functionality. Each step is described in detail below.

\subsection{Knowledge Base Generation}
\subsubsection{The Requirements Specification Knowledge Base (\(K_R\))}
As depicted by the step marked as `A1’ in Figure~\ref{fig:detection}, a knowledge base is constructed exclusively from the content of the requirements specification document. The individual requirements are chunked one by one to preserve their granularity. The remainder of the document, which typically includes contextual sections such as the scope, system overview, definitions, constraints, references, and appendices, is segmented into fixed-length chunks of 400 tokens, with an overlap of 20 tokens between consecutive chunks. Each chunk is then vectorized using OpenAI’s \texttt{text-embedding-3-small} model~\cite{openai2024embedding} to generate embeddings that capture its semantic meaning. This encoder model was selected due to its strong performance on semantic textual similarity benchmarks and its cost-effectiveness for large-scale embedding generation. The resulting vectors are stored in Pinecone~\cite{pinecone2025}, a scalable vector database, to enable efficient similarity search and retrieval. This collection of vectors constitutes the requirements specification knowledge base, \(K_R\).

\subsubsection{The Domain Knowledge Bases (\(K_N\), \(K_I\), and \(K_E\))}
As shown in step `A2' of Figure~\ref{fig:detection}, domain knowledge bases 
are constructed using WikiDoMiner~\cite{wikidominer}, an automated tool for 
retrieving domain-relevant documents from Wikipedia. WikiDoMiner accepts the 
requirements specification document as input, scores candidate keywords using 
TF-IDF, and selects the top-$K$ terms to query Wikipedia. The retrieval scope 
is controlled by a configurable depth parameter:
\begin{itemize}
    \item Depth 0 (\textit{Novice}): Retrieves only the Wikipedia articles that directly match the extracted keywords. The retrieved documents are used to populate the \textit{novice} domain knowledge base, \(K_N\).
    \item Depth 1 (\textit{Intermediate}): Expands the scope by including articles from the categories of the keyword-matched articles. These documents populate the \textit{intermediate} domain knowledge base, \(K_I\).
    \item Depth 2 (\textit{Expert}): Further extends the retrieval to include articles from subcategories of the keyword-matched articles, yielding the broadest coverage of domain knowledge. These documents populate the \textit{expert} domain knowledge base, \(K_E\).
\end{itemize}

Each document set is segmented into fixed-length chunks of 400 tokens with an overlap of 20 tokens. The chunks are encoded into embeddings using OpenAI’s \texttt{text-embedding-3-small} model \cite{openai2024embedding}. The resulting vectors for each depth are stored in separate Pinecone namespaces \cite{pinecone2025}, forming the three domain knowledge bases (\(K_N\), \(K_I\), and \(K_E\)). 

Wikipedia is used as a practical source to simulate different levels of domain knowledge. The intent is not to model real stakeholder expertise, but to capture relative differences in their domain knowledge. Wikipedia’s category structure allows the retrieval scope to be expanded in a systematic and repeatable way by varying traversal depth, where increasing depth corresponds to broader conceptual coverage and exposure to a larger set of related domain concepts \cite{wikidominer}. Such graph-based expansion has been used in prior work to approximate increasing domain context and conceptual breadth \cite{chi1981categorization,navigli2012babelnet}. In addition, Wikipedia provides reasonably consistent coverage across most domains \cite{milne2006mining}. While sources such as textbooks and technical manuals may better reflect expert-level knowledge, their use would require manual curation, would vary widely in availability across different domains, and would often involve licensing restrictions. These factors would make it difficult to apply the same retrieval process consistently across domains and are, therefore, outside the scope of this work.

\subsection{Elucidation Question Generation (EQGen)}
\label{EQGen}
Elucidation Question Generation (EQGen), depicted by the step marked as `B' in Figure~\ref{fig:detection}, leverages an LLM to identify potentially ambiguous terms within each requirement and generate a corresponding Elucidation Question (EQ) for each. An ambiguous term is a word or phrase that could reasonably be interpreted in more than one way by stakeholders reading the requirement. For each such term, an EQ is a clarification question that helps narrow down the meaning of the ambiguous term so that all stakeholders would interpret the requirement consistently. EQs are scoped to requirement-level ambiguities that affect the intent or scope of the system --- proper nouns, standardized interface names, product names, and specific technical standards are excluded, as are questions about implementation details, document versions, or operational procedures.

The output of EQGen consists of a list of EQs for each requirement, each EQ associated with an identified ambiguous term. If no ambiguous terms are detected within a particular requirement from the requirements specification document, the requirement is labeled as ``unambiguous,'' and no EQs are generated. It is important to note that this step detects linguistic indicators of potential ambiguity, but does not in itself confirm pragmatic ambiguity. The subsequent steps determine whether differing interpretations truly emerge under varying domain knowledge, thereby establishing pragmatic ambiguity.

\begin{figure}[h] 
    \centering
    \includegraphics[width=1\columnwidth]{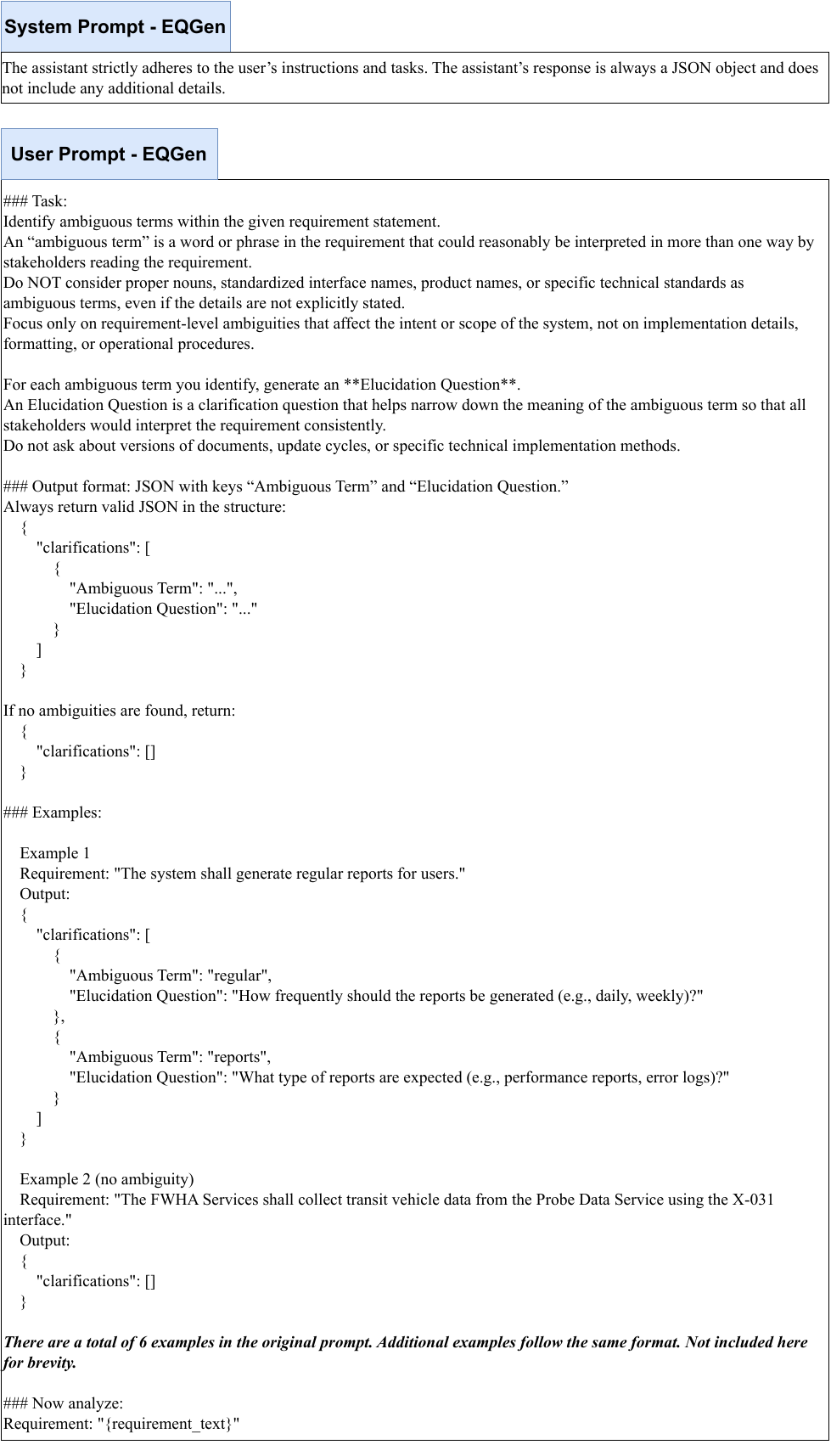} 
    \caption{Prompt used for Elucidation Question Generation}
    \label{fig:prompt_EQ}
\end{figure}

The final prompt used for EQGen was developed through an iterative prompt engineering process, inspired by the principles on prompt design \cite{phoenix2024promptengineering}. Our prompt development process involved testing variations in system instructions, example coverage, and output format. The final prompt is shown in Figure~\ref{fig:prompt_EQ}.

\subsection{Searching in the Requirements Specification Knowledge Base (\(K_R\))}
\begin{figure}[h] 
    \centering
    \includegraphics[width=0.7\columnwidth]{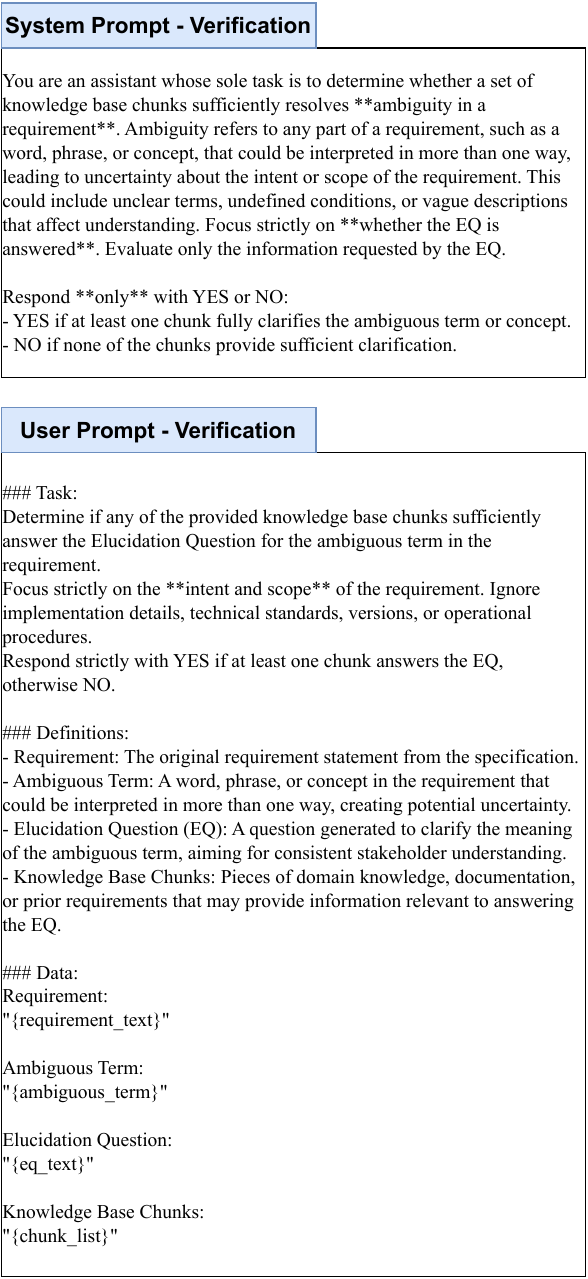} 
    \caption{Prompt Used for Verifying Whether Chunks Retrieved from \(K_R\) Answer an EQ}
    \label{fig:verify}
\end{figure}

This phase is shown by the steps marked as `C1’-`C3’ in Figure~\ref{fig:detection}. It involves querying \(K_R\) to identify relevant information that may address the generated EQs. For each EQ, the system searches \(K_R\) and retrieves the top 5 most similar text chunks, ranked by their cosine similarity score to the embedding of the requirement and its EQ (step marked as `C1’ in Figure~\ref{fig:detection}). The system employs the same LLM used in the EQGen phase to verify whether the information in the retrieved chunk effectively answers the EQ (steps marked as `C2' in Figure~\ref{fig:detection}). The prompt used for verification is shown in Figure~\ref{fig:verify}. If the retrieved information answers all the EQs (step marked as `C3’ in Figure~\ref{fig:detection}), the corresponding requirement is labeled as pragmatically unambiguous. If no valid answer is found within \(K_R\), the system proceeds to consult the domain knowledge bases (\(K_N\), \(K_I\), and \(K_E\)).

\subsection{Searching in the Domain Knowledge Bases (\(K_N\), \(K_I\), and \(K_E\))}
\label{domain_search}
This phase, depicted by the steps marked as `D1’- `D3’ in Figure~\ref{fig:detection}, focuses on assessing the consistency of interpretations derived from \(K_N\), \(K_I\), and \(K_E\). In this context, ``interpretations'' refer to the chunks of text retrieved from a domain knowledge base that are most similar to the requirement and its corresponding EQ. Each retrieved chunk represents how a stakeholder with the corresponding level of domain knowledge (\textit{novice}, \textit{intermediate}, or \textit{expert}) might understand or resolve the ambiguity inherent in the requirement.
When no answer is found in \(K_R\), the system queries the domain knowledge bases (\(K_N\), \(K_I\), and \(K_E\)). It retrieves the top 3 most relevant vectors from each domain knowledge base, based on their cosine similarity to the requirement and EQ’s embedding (step marked as `D1’ in Figure~\ref{fig:detection}).
To determine whether a requirement is pragmatically unambiguous, the system evaluates the consistency of interpretations retrieved from \(K_N\), \(K_I\), and \(K_E\). Specifically, it constructs groups by combining the 3 retrieved chunks from each domain knowledge base in all possible combinations, each group containing one chunk from each domain knowledge base. This results in a total of $27 \,(3^3)$ groups, each group containing $3$ chunks. The pairwise cosine similarity scores between the embeddings of the chunks within each group are computed (step marked as `D2’ in Figure~\ref{fig:detection}). If any such group exhibits pairwise similarity scores that all exceed a pre-determined threshold \(T\) (see Section~\ref{threshold}), the requirement is considered pragmatically unambiguous, indicating that the term is interpreted consistently across different domain knowledge levels (step marked as `D3’ in Figure~\ref{fig:detection}). Conversely, if none of the groups meet this threshold, the requirement is marked as pragmatically ambiguous, indicating that interpretations differ across the various levels of domain knowledge.

\subsection{Pragmatic Ambiguity Resolution}
\label{sec:resolution}
\begin{figure}[h] 
    \centering
    \includegraphics[width=0.9\columnwidth]{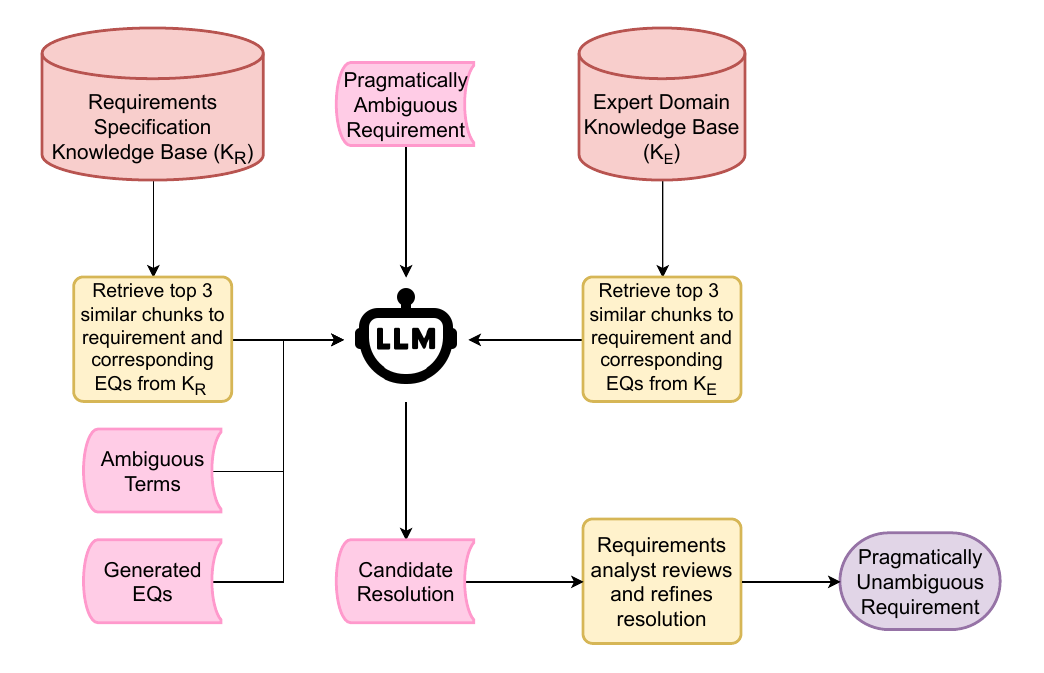} 
    \caption{Proposed Approach for Pragmatic Ambiguity Resolution}
    \label{fig:resolution}
\end{figure}

\begin{figure}[h] 
    \centering
    \includegraphics[width=0.7\columnwidth]{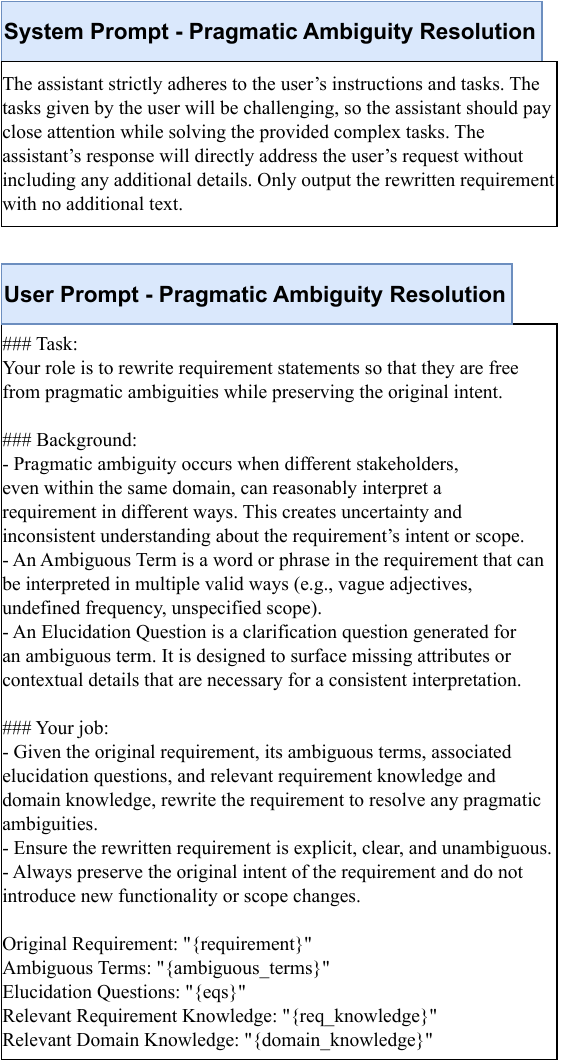} 
    \caption{Prompt Used for Generation of Candidate Requirements}
    \label{fig:prompt_resolution}
\end{figure}

The resolution phase, illustrated in Figure~\ref{fig:resolution}, generates candidate resolutions for those requirements flagged as pragmatically ambiguous. Each pragmatically ambiguous requirement, along with its associated EQs, is queried against both \(K_R\) and \(K_E\) to retrieve relevant contextual information. Candidate resolutions are then generated using an LLM prompted with a carefully designed instruction, shown in Figure~\ref{fig:prompt_resolution}. The inputs provided to the prompt include:
\begin{itemize}
    \item Original Requirement – the raw requirement flagged as pragmatically ambiguous.
    \item Ambiguous Terms – words or phrases in the requirement that could be interpreted in multiple ways.
    \item EQs – clarification questions generated for each ambiguous term.
    \item Requirement Knowledge – top 3 retrieved chunks (ranked by their cosine similarity score to the embedding of the requirement and its EQ) from \(K_R\).
    \item Domain Knowledge – top 3 retrieved chunks (ranked by their cosine similarity score to the embedding of the requirement and its EQ) from \(K_E\).
\end{itemize}

The prompt was developed through an iterative prompt engineering process, inspired by established principles of prompt design \cite{phoenix2024promptengineering}. During development, variations in system instructions were tested to ensure that the LLM consistently produced resolutions that preserved the original intent.

The candidate resolutions must then be reviewed by an RA, who verifies whether the rewritten requirement accurately captures the intended meaning and ensures that disambiguation does not introduce inconsistencies or alter the requirement’s scope. It is important to note that while our approach leverages domain knowledge to resolve pragmatic ambiguity, it does not guarantee alignment with the intended system functionality; final validation by the RA remains essential.

\section{Experimental Setup}
\label{experimental_setup}
In this section, we provide a detailed description of the experimental setup. All experiments were conducted using four LLMs: GPT-4o-mini \cite{openai2024gpt4omini}, Llama-3.1-8B\footnote{https://huggingface.co/meta-llama/Llama-3.1-8B-Instruct}, Mistral-7B \footnote{https://huggingface.co/mistralai/Mistral-7B-Instruct-v0.3}, and Qwen2.5-7B\footnote{https://huggingface.co/Qwen/Qwen2.5-7B-Instruct}. GPT-4o-mini is a widely used commercial model known for strong performance on general NLP tasks. Mistral-7B, Llama-8B, and Qwen2.5-7B are open-weight models with competitive performance and are accessible for reproducible research. The selected models vary in architecture and training data, enabling us to assess the robustness of our approach across both closed models and open models in the 7–8B parameter range. The link to the ground truth datasets, code, and intermediate outputs generated in the experiments can be found here.\footnote{\url{https://github.com/pavithranair/Pragmatica}}
\subsection{Dataset Creation}
The first document used in this study is ``Clarus Weather System Design,'' a high-level system requirements specification issued by the U.S. Department of Transportation, Federal Highway Administration, in 2005, which includes 140 requirements. The second document is ``Vehicle Infrastructure Integration (VII) Data Use Analysis and Processing (DUAP),'' issued by the Michigan Department of Transportation (MDOT) in 2007, containing 148 requirements. 

The Clarus Weather System Design specification provides a repository of high-level requirements governing the design of the Clarus system, an initiative aimed at organizing and enhancing environmental and road condition observation capabilities. The VII Data Use Analysis and Processing document outlines the requirements for Michigan’s vehicle infrastructure integration DUAP System, which investigates how new vehicle infrastructure integration data impacts safety, traffic operations and management, asset management, winter operations, and transportation planning. These documents were selected because they offer diverse contexts and complex requirements, making them ideal for evaluating the proposed approach.

\begin{figure}[h] 
    \centering \includegraphics[width=0.9\columnwidth]{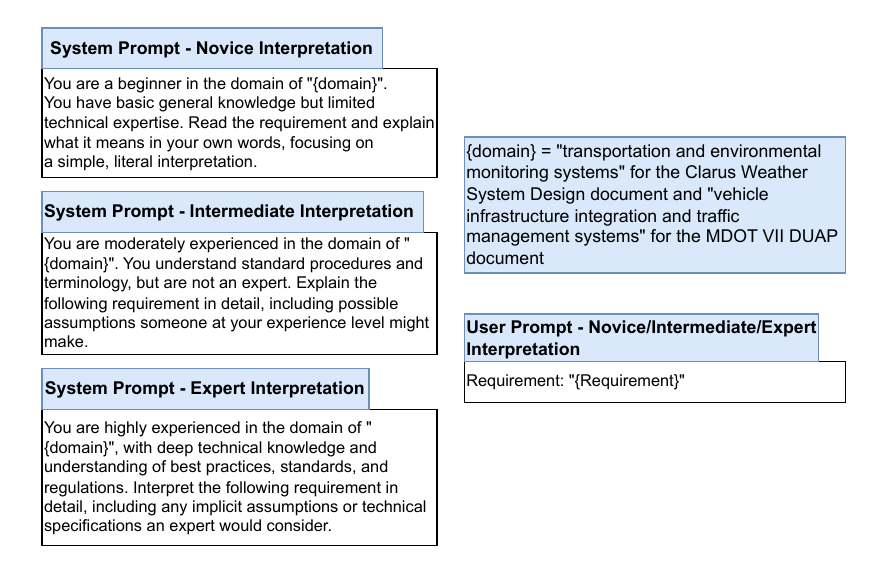} 
    \caption{Prompts Used to Generate Novice-, Intermediate-, and Expert-level Interpretations of Requirements to Support the Annotation of Pragmatic Ambiguities}
    \label{fig:interpretations}
\end{figure}

From the two documents, requirements exhibiting pragmatic ambiguities were 
manually identified and tagged. Specifically, 37 requirements from the Clarus 
document and 39 requirements from the VII DUAP document were marked as 
pragmatically ambiguous. Evaluators were selected based on their demonstrated 
expertise in transportation systems requirements, with a minimum of five years 
of industrial research experience in the application domain of the evaluated 
documents. Two such domain experts, each with 5--7 years of industrial 
experience as researchers in transportation systems and environmental/road 
condition monitoring, independently assigned binary labels to indicate whether 
each requirement was pragmatically ambiguous. Prior to annotation, the domain 
experts were instructed to carefully review the requirements specification 
documents to ensure that any pragmatic ambiguity marked could not be reasonably 
inferred or resolved from the specification document itself. To support their 
assessment, novice, intermediate, and expert-level interpretations of each 
requirement, generated using GPT-4~\cite{openai2023gpt4}, were provided; 
however, the experts were instructed to use these interpretations only as 
guidance and rely primarily on their own judgment. The prompts used to generate 
these interpretations are shown in Figure~\ref{fig:interpretations}. Cohen's 
Kappa was computed to assess inter-annotator consistency, yielding a score of 
0.89. In cases of disagreement, a senior domain expert with 15 years of 
experience as an industrial researcher in transportation systems reviewed the 
requirement and provided the final label. Each participant spent approximately 
three working days on this task, averaging two hours per day. None of the domain 
experts who participated in the annotation task, and later in the evaluation of 
the candidate resolutions (see Section~\ref{resolution_evaluation}), are 
co-authors of this paper. Their participation was voluntary, and no monetary 
compensation was provided. The experts were informed about the purpose of the 
study and contributed their time based on their professional interest in the 
research topic. A stratified split was performed, with 60\% of the labelled 
dataset used for threshold calibration (see Section~\ref{threshold}) and the 
remaining 40\% held out for reporting the final pragmatic ambiguity detection 
results.

WikiDoMiner~\cite{wikidominer} was used to construct \(K_N\), \(K_I\), and 
\(K_E\) by extracting the top-$K$ keywords from each requirements specification 
using TF-IDF scoring and querying Wikipedia at three depth levels. For the 
Clarus document, \(K_N\) was populated with 99 Wikipedia pages, \(K_I\) with 
5,385 pages, and \(K_E\) with 31,774 pages. For the VII DUAP document, \(K_N\) 
included 161 pages, \(K_I\) included 20,188 pages, and \(K_E\) included 45,722 
pages. For novice-level retrieval (depth = 0), only pages matching the extracted 
keywords were included. Intermediate-level retrieval (depth = 1) expanded to 
include pages from categories associated with the keyword-matched articles, and 
expert-level retrieval (depth = 2) further incorporated pages from subcategories 
of these articles. For example, Clarus keywords such as ``weather forecasting'' 
and ``road condition observation'' were used to retrieve corresponding Wikipedia 
articles; associated categories included ``Environmental Monitoring'' and 
``Transport,'' while subcategories extended to ``Environmental Impact 
Assessment'' and ``Transport Infrastructure.''

\subsection{Evaluation Metrics}
We employ the following evaluation metrics to assess the performance of the approach:
\subsubsection{Detection of Pragmatic Ambiguity}
We compute accuracy, precision, recall, and the F1 and F2 scores for all experiments. Given the potential consequences of failing to identify pragmatic ambiguities during requirement implementation, it is crucial to minimize missed detections (false negatives). The F2 score places greater emphasis on recall, which is critical in ensuring that pragmatic ambiguities are identified and addressed.

\subsubsection{Candidate Resolutions}
\label{resolution_evaluation}
The ISO/IEC/IEEE 29148:2018 standard \cite{ISO29148_2018} outlines essential quality characteristics for individual requirements, highlighting the importance of lack of ambiguity, clarity, and consistency. To align with these principles and assess the quality of candidate resolutions, we employ three primary evaluation metrics:

\begin{itemize}
    \item \textit{Relevance}: This metric assesses how well the suggested resolution addresses the identified pragmatic ambiguity. A resolution is deemed relevant if it directly addresses the specific pragmatic ambiguity.
    \item \textit{Clarity}: \textit{Clarity} measures the ease with which the resolution can be understood by stakeholders. A resolution is considered clear if it is easily understandable and presents the solution in a straightforward manner.
    \item \textit{Consistency}: \textit{Consistency} evaluates how well the resolution aligns with the overall context of the requirements specification document. A consistent resolution does not introduce contradictions when integrated with other parts of the requirements specification document.
\end{itemize}

The same two domain experts employed in the initial annotation of pragmatic ambiguities independently rated each candidate resolution on these metrics using a 3-point scale: 3 if the resolution fully met the criterion, 2 if it partially met the criterion, and 1 if it failed to meet the criterion. The evaluators were provided with the pragmatically ambiguous requirement, the detected ambiguous terms, the generated EQs, the requirements specification document, and the candidate resolution. Inter-annotator agreement was calculated using weighted Cohen’s Kappa across all metrics, yielding an overall agreement score of 0.77. 

\subsection{Threshold Calibration}
\label{threshold}
A critical component of the pragmatic ambiguity detection phase is determining the similarity threshold \(T\), which governs whether interpretations retrieved from \(K_N\), \(K_I\), and \(K_E\) are considered consistent (see Section~\ref{domain_search}). If all pairwise cosine similarity scores within a group of interpretations exceed \(T\), the requirement is classified as pragmatically unambiguous. 

To identify an appropriate value for \(T\), we performed stratified 5-fold cross-validation on the training split (60\% of the dataset). For each fold, we swept threshold values in the range 0.70–0.98, incrementing by 0.01, and computed the F2 score. The optimal threshold for each fold was selected as the value of \(T\) that maximized the F2 score. The final threshold was then computed as the mean of the fold-level optimal thresholds. This process was repeated independently for each of the four LLMs used in our experiments. The models and their corresponding calibrated thresholds are reported in Table~\ref{tab:thresholds}.
\setlength{\tabcolsep}{2pt} 
\begin{table}[h]
\centering
\caption{Calibrated Thresholds for LLMs Used in Pragmatic Ambiguity Detection}
\label{tab:thresholds}
\begin{tabular}{lcccc}
\toprule
\textbf{Model} & GPT-4o-mini & Llama-3.1-8B & Mistral-7B & Qwen2.5-7B \\
\midrule
\textbf{Threshold} & 0.87 & 0.85 & 0.82 & 0.86 \\
\bottomrule
\end{tabular}
\end{table}

\section{Results and Analysis}
\label{results}
\subsection{Pragmatic Ambiguity Detection (RQ\textsubscript{1})}
\setlength{\tabcolsep}{3pt} 

\begin{table}[b]
\centering
\caption{Pragmatic Ambiguity Detection Results for the Clarus Weather System Design and MDOT VII DUAP Documents}
\label{tab:detection_results}
\begin{tabular}{lccccc}
\toprule
\textbf{Model} & \makecell{\textbf{Macro} \\ \textbf{Accuracy}} & \makecell{\textbf{Macro} \\ \textbf{Precision}} & \makecell{\textbf{Macro} \\ \textbf{Recall}} & \makecell{\textbf{Macro} \\ \textbf{F1}} & \makecell{\textbf{Macro} \\ \textbf{F2}} \\
\midrule
\multicolumn{6}{c}{\textbf{Clarus Weather System Design}} \\
\midrule
\textbf{GPT-4o-mini}    & \textbf{0.75} & \textbf{0.73} & \textbf{0.75} & \textbf{0.73} & \textbf{0.74} \\
Llama-3.1-8B   & 0.64 & 0.6  & 0.6  & 0.6  & 0.6  \\
Mistral-7B     & 0.66 & 0.63 & 0.64 & 0.63 & 0.64 \\
Qwen2.5-7B     & 0.68 & 0.63 & 0.6  & 0.61 & 0.6  \\
\midrule
\multicolumn{6}{c}{\textbf{MDOT VII Data Use Analysis and Processing}} \\
\midrule
\textbf{GPT-4o-mini}   & \textbf{0.8}  & \textbf{0.76} & \textbf{0.76} & \textbf{0.76} & \textbf{0.76} \\
Llama-3.1-8B   & 0.65 & 0.6  & 0.61 & 0.6  & 0.6  \\
Mistral-7B     & 0.67 & 0.61 & 0.62 & 0.61 & 0.62 \\
\textbf{Qwen-2.5-7B}    & \textbf{0.8}  & \textbf{0.76} & \textbf{0.76} & \textbf{0.76} & \textbf{0.76} \\
\bottomrule
\end{tabular}
\end{table}

\begin{table*}[h]
\centering
\caption{Analysis of Pragmatic Ambiguity Detection Results}
\label{tab:detection_analysis}
\begin{tabular}{|p{4.5cm}|p{1.5cm}|p{6.5cm}|p{1.1cm}|p{3cm}|}
\hline
\textbf{Requirement} & \textbf{Ambiguous Term} & \textbf{Generated Elucidation Question} & \textbf{Result} & \textbf{Explanation} \\
\hline
The DUAP Output Services shall be able to publish traffic data to TMS using the TMS interface format. 
& TMS interface format 
& Can you provide more details about the specific format used for publishing traffic data to TMS, such as the data structure and any required standards or guidelines? 
& False \newline Positive 
& Dependence on \newline external \newline agreement/document \\
\hline
The information browsers shall provide a means for users to select between Metric and English units of measurement for display. 
& shall provide 
& Is it necessary for the system to provide a built-in UI control for unit selection, or can users make this choice through other established methods within the browser (e.g., settings)? 
& False \newline Positive 
& Irrelevant EQ \\
\hline
The DUAP System shall allow data to be accessed as soon as it is cached. 
& cached 
& What defines the `cache' in this context? Is it a buffer, temporary storage, or a specific layer in the system architecture? 
& False \newline Positive 
& Trivial ambiguity \\
\hline
The DUAP Computational Services shall be able to derive pothole severity from probe vehicle data. 
& N/A 
& N/A 
& False \newline Negative 
& No ambiguous terms were identified \\
\hline
The Clarus system shall be able to detect data submission errors. 
& data submission errors 
& What specific types of data submission errors should the system be able to detect (e.g., formatting errors, missing fields, validation errors)? 
& False \newline Negative 
& Ambiguous term correctly identified, but domain knowledge bases provided vague yet similar interpretations \\
\hline
\end{tabular}
\end{table*}
The pragmatic ambiguity detection results for Clarus Weather System Design and MDOT VII DUAP requirements specification
documents are presented in Table~\ref{tab:detection_results}. For Clarus, GPT-4o-mini achieved the highest F2 score (0.74). Mistral-7B and Qwen2.5-7B showed moderate performance. For the MDOT VII DUAP document, GPT-4o-mini and Qwen2.5-7B performed similarly, achieving the highest F2 scores (0.76). 

False positives occur when requirements reference external documents or agreements not included in the domain knowledge bases. Trivial ambiguities, where the impact of detected ambiguous terms on overall requirement clarity is minimal, also contribute. False negatives typically occur when ambiguous terms are missed during EQGen or when domain knowledge base interpretations are vague. In such cases, the domain interpretations may fail to provide clear answers to the EQ, yet are sufficiently similar to evade detection. Table~\ref{tab:detection_analysis} provides illustrative examples of these error cases.

\begin{framed}
\noindent\textbf{Answer to RQ\textsubscript{1}}\\[3pt]
Overall, these results indicate that the proposed approach can effectively detect pragmatic ambiguities, while the observed error cases highlight limitations related to EQGen and the availability of domain knowledge bases with sufficient variation to surface interpretation differences.
\end{framed}

\subsection{Pragmatic Ambiguity Resolution (RQ\textsubscript{2})}
The pragmatic ambiguity resolution results are presented in Table~\ref{tab:resolution_results}. Evaluation was performed on the candidate resolutions generated for all requirements labelled as pragmatically ambiguous in the ground truth datasets. For Clarus, Mistral-7B achieved the highest scores for \textit{Clarity} (2.95) and \textit{Consistency} (2.93). GPT-4o-mini achieved the highest \textit{Relevance} score (2.53), showing strong performance in directly addressing the identified ambiguities. Llama-3.1-8B and Qwen2.5-7B showed slightly lower performance across the three metrics. For MDOT, GPT-4o-mini achieved the highest \textit{Relevance} score (2.81). Mistral-7B achieved the highest \textit{Clarity} (2.91) and \textit{Consistency} (3.0) scores. Llama-3.1-8B and Qwen2.5-7B showed moderate performance across all three metrics.

A few challenges were observed during the generation of candidate resolutions. Some resolutions introduced unnecessary complexity or speculative language that was not present in the original requirement. In certain cases, a resolution successfully addressed the existing ambiguity but inadvertently changed the original meaning or introduced new ambiguities. Other resolutions only partially resolved the underlying pragmatic ambiguity, leaving residual uncertainty. Table~\ref{tab:resolution_analysis} provides illustrative examples of these issues.

\begin{table}[t]
\centering
\caption{Human Evaluation Scores for Candidate Resolutions on the Clarus Weather System Design and MDOT VII DUAP Documents. Values are mean $\pm$ standard deviation.}
\label{tab:resolution_results}
\begin{tabular}{lccc}
\toprule
\textbf{Model} & \textbf{Relevance} & \textbf{Clarity} & \textbf{Consistency} \\
\midrule
\multicolumn{4}{c}{\textbf{Clarus Weather System Design}} \\
\midrule
GPT-4o-mini & \textbf{2.53 $\pm$ 0.72} & 2.82 $\pm$ 0.45 & 2.88 $\pm$ 0.33 \\
Llama-3.1-8B         & 2.46 $\pm$ 0.81 & 2.91 $\pm$ 0.37 & 2.85 $\pm$ 0.48 \\
\textbf{Mistral-7B}            & 2.47 $\pm$ 0.64 & \textbf{2.95 $\pm$ 0.23} & \textbf{2.93 $\pm$ 0.25} \\
Qwen2.5-7B            & 2.41 $\pm$ 0.73 & 2.73 $\pm$ 0.44 & 2.91 $\pm$ 0.29 \\
\midrule
\multicolumn{4}{c}{\textbf{MDOT VII Data Use Analysis and Processing}} \\
\midrule
GPT-4o-mini & \textbf{2.81 $\pm$ 0.75} & 2.77 $\pm$ 0.53 & 2.85 $\pm$ 0.51 \\
Llama-3.1-8B         & 2.72 $\pm$ 0.60 & 2.68 $\pm$ 0.61 & 2.78 $\pm$ 0.57 \\
\textbf{Mistral-7B}            & 2.71 $\pm$ 0.56 & \textbf{2.91 $\pm$ 0.33} & \textbf{3.00 $\pm$ 0.00} \\
Qwen2.5-7B            & 2.71 $\pm$ 0.56 & 2.83 $\pm$ 0.46 & 2.91 $\pm$ 0.36 \\
\bottomrule
\end{tabular}
\end{table}

\begin{table*}[h]
\centering
\renewcommand{\arraystretch}{0.9} 
\caption{Analysis of Pragmatic Ambiguity Resolution Results}
\label{tab:resolution_analysis}
\begin{tabular}{|p{2.5cm}|p{1.5cm}|p{3.5cm}|p{7.7cm}|p{1.5cm}|}
\hline
\textbf{Pragmatically\newline Ambiguous \newline Requirement} & \textbf{Ambiguous Term(s)} & \textbf{Generated Elucidation \newline Question(s)} & \textbf{Candidate Resolution (`...' indicates text has been shortened)} & \textbf{Explan\-ation} \\
\hline
The DUAP Computational Services shall be able to derive pothole location from probe vehicle data. & pothole \newline location & What specific criteria or measurements will define a 'pothole' for the purpose of this system? & The DUAP Computational Services shall be able to derive pothole locations defined as depressions in the road surface, where the underlying soil is weakened by moisture and traffic, from probe vehicle data, using specific criteria that measure size, depth, and surface conditions according to standard road assessment methodologies. & Partial \newline resolution \\
\hline
The DUAP System shall be able to calculate measures of effectiveness. & measures of \newline effectiveness & What specific metrics or indicators should the DUAP System calculate, and how will these be defined? & The DUAP System shall calculate the following specific measures of effectiveness: Daylight Autonomy (DA)...100 lux as UDI-s and those above 3000 lux as UDI-e. & Misaligned with \newline original intent \\
\hline
The DUAP System shall archive cached data. & archive & What specific criteria and retention policies should be applied to the archived cached data? & The DUAP System shall archive cached data according to specified criteria and retention policies, ensuring compliance with relevant legal and business data archival requirements. The cached data, both raw and derived, shall be retained for a minimum of one week and then subjected to defined actions at the end of the retention period, including secure deletion or further archiving as per established guidelines. & Introduces new \newline ambiguity and \newline unnecessary complexity \\
\hline
\end{tabular}
\end{table*}

\begin{framed}
\noindent\textbf{Answer to RQ\textsubscript{2}}\\[3pt]
Overall, the results indicate that the proposed approach can generate candidate resolutions that are relevant, clear, and consistent with the intended system functionality, while highlighting the need for RA validation to avoid unnecessary complexity or unintended changes in meaning.
\end{framed}

\section{Threats to Validity}
\label{threats}
The threats and limitations of our approach can be categorized into three main areas: internal, construct, and external validity.

\subsection{Internal Validity}
Our approach models only three levels of domain knowledge, which may not capture the full range of possible stakeholder interpretations that could arise in practice. The retrieval parameters were set experimentally to balance performance and computational efficiency. Specifically, we used the top 5 results from \(K_R\) for EQGen (see Section~\ref{EQGen}), the top 3 from each domain knowledge base (see Section~\ref{domain_search}), and the top 3 from both \(K_R\) and \(K_E\) for resolution generation (see Section~\ref{sec:resolution}). These parameter choices may affect detection outcomes. In the ground truth dataset labelling, domain experts may not have been fully able to capture where a stakeholder with novice or intermediate domain knowledge might interpret a requirement differently, even with GPT-4-generated interpretations at multiple domain expertise levels for guidance. Furthermore, providing GPT-4-generated interpretations during annotation may have introduced bias, as experts could have been anchored to the 
perspectives presented by the model rather than independently arriving 
at their own assessments.

\subsection{Construct Validity}
Candidate resolution quality is assessed using human evaluations on \textit{Relevance}, \textit{Clarity}, and \textit{Consistency}. These metrics inherently rely on subjective judgment by domain experts. To mitigate this, we employed multiple experts, computed inter-annotator agreement, and provided structured evaluation guidelines for the evaluation of pragmatic ambiguity resolution. Even so, some dimensions of pragmatic ambiguity may not be fully captured by these three metrics.


\subsection{External Validity}
The applicability of our findings may be influenced by the specific 
requirements specification documents and domain knowledge bases used in 
this study. We evaluated two distinct transportation-related documents. Pragmatic 
ambiguity detection and resolution outcomes could vary in other domains 
or contexts. To reduce this threat, the two documents were selected from 
different sub-domains within transportation to introduce some variation 
in terminology and domain context. Future studies should validate the 
approach on documents from domains such as healthcare, finance, and 
industrial software, and on larger datasets to assess generalizability 
and reduce the risk of threshold overfitting.

\section{Conclusion}
\label{conclusion}
We present a RAG-based framework for detecting and resolving pragmatic 
ambiguities in NLRs by simulating stakeholder interpretations across 
novice, intermediate, and expert domain knowledge bases. Requirements 
whose interpretations diverge across expertise levels are flagged as 
pragmatically ambiguous, and candidate resolutions are proposed for RA 
validation.

To assess the effectiveness of our approach, we use four models: GPT-4o-mini, Llama-3.1-8B, Mistral-7B, and Qwen2.5-7B, on a dataset comprising two requirements specification documents. GPT-4o-mini outperformed the other models in detecting pragmatic ambiguities and achieved the highest human evaluation scores for \textit{Relevance} in pragmatic ambiguity resolution. Mistral-7B achieved the highest human evaluation scores for \textit{Clarity} and \textit{Consistency}. Overall, the results indicate that the proposed approach is effective in detecting pragmatic ambiguities and generating candidate resolutions that are judged by domain experts to be relevant, clear, and consistent with the intended system functionality.

For future work, we plan to extend the framework by integrating stakeholder feedback loops, allowing the system to refine candidate resolutions based on the preferences of multiple stakeholders. We also aim to model conflicting priorities among stakeholders, so that disambiguation can account for trade-offs between competing goals, such as safety, cost, or performance. Additionally, we will explore how organizational context influences requirement interpretation, enabling the framework to adapt its recommendations to different procedural or institutional norms. 

\bibliographystyle{ACM-Reference-Format}
\bibliography{acmart}

\end{document}